\documentclass[showpacs,prb,aps,twocolumn]{revtex4-1}

\pdfoutput=1

\usepackage{amsmath,amsfonts,amssymb,bm,graphicx,hyperref,color,units}

\newcommand{\Pa}{\mathcal{P}}
\newcommand{\T}{\mathcal{T}}

\newcommand{\OO}{\mathcal{O}}
\newcommand{\Com}{\mathbb{C}}

\newcommand{\bra}[1]{\langle #1|}
\newcommand{\ket}[1]{|#1\rangle}
\newcommand{\Tr} {{\text T}{\text r}}

\begin{document}

\title{Timescales, dynamical phase transitions and $3$rd order phase transitions
in the $1$d anisotropic $XY$ model}
\author{James M. Hickey}
\affiliation{School of Physics and Astronomy, University of Nottingham,
Nottingham, NG7 2RD, United Kingdom}
\date{\today}

\begin{abstract}
Recent research into time-integrated observables has revealed a special class of
states which capture the singular features of the generating functions of those
observables, as estimated by full counting statistics (FCS).  In this work we
extend the results of [Phys. Rev. B \textbf{87} 184303 (2013)] to the $1$d
anisotropic $XY$-model and find a set of FCS critical points
associated with the time-integrated transverse magnetization and anisotropy. We
show dynamical phase
transitions (DPTs) as defined in [Phys. Rev. Lett. \textbf{110} 135704 (2013)]
do not emerge on quenching the states associated with the time-integrated
anisotropy across FCS critical points.
We also study the timescales required to prepare the associated special states of the transverse magnetization using appropriate Markovian baths and find they are
independent of the number of spins in the chain. However the probability to
evolve to such a state decreases drastically with increasing system size due to
dissipation.
Thus when preparing such states it is preferable to use few body systems and to
reach the thermodynamic limit it is necessary to use an ancillary system in
conjunction with the Markovian baths.

\end{abstract}

\maketitle 

\section{Introduction}
\label{sec:Intro}
The collective behaviour of many elementary single particles combined in
an interacting system can be very rich and differ greatly from the physics
governing each individual particle~\cite{Maxwell1875,Peliti2011}.  One of
the most remarkable phenomena in nature arising from collective behaviour is
the phase transition, whereby tuning a physical field the system may undergo
a singular change in its static properties.   This phenomenom is ubiquitous in nature and so the theory of phase transitions naturally extends into quantum systems with the paradagmatic
model of a $2$nd order quantum phase transition being the $1$d anisotropic $XY$ model in a transverse field.  By tuning the transverse field the ground state undergoes a singular change from a ferromagnetic state to a
paramagnetic state~\cite{Sachdev2011}.

However focussing on equilibrium properties does not necessarily capture
the full dynamical behaviour of a complex many-body system~\cite{Barrat2004}. 
Recent studies have found that consideration of purely dynamical observables, more specifically
time-integrated observles~\cite{Ruelle2004,Lecomte2007,Garrahan2007},  provide
insights into the dynamics of many-body systems.  Using full counting statistics
(FCS)~\cite{Lecomte2007,Garrahan2007,Levitov1993,*Levitov1996,Nazarov2003,*Nazarov2003b,Pilgram2003,Flindt2008,Esposito2009,Flindt2009}
the moment generating function (MGF) of these observables can be calculated and
is treated analogously to a partition sum.  Moreover the counting field `$s$',
that is the field conjugate to the time-integrated observable, is treated like a full thermodynamic variable~\cite{Hedges2009,Pitard2011,*Speck2012}.
Within this formalism singular features in the long-time limit of the cumulant
generating function (CGF) mark phase transitions in the FCS.  

In a recent paper~\cite{Hickey2013} we focussed on the time-integrated
transverse magnetization in the ground state of an associated model, the transverse field
Ising model (TFIM), using this approach.
We uncovered a whole curve of $2$nd order FCS phase transitions of which the static quantum
critical points were the endpoints.  We also found that a special class of
states exist which capture these FCS singular points in an analogous manner to
the ground state.  By considering this critical curve
as a quantum critical line in this extended class of states these FCS singular points corresponded to a
$3$rd order static phase transition which was probed by tuning the field $s$. 
These special states were identified as eigenstates of a non-Hermitian operator which forms
the MGF.  Relating this non-Hermitian operator to the effective Hamiltonian of
an appropriate open quantum system it was shown that the state could be prepared
via the no jump evolution of this open system~\cite{Hickey2013,Plenio1998}.  A
second approach to dynamics and quantum nonequilibrium focusses on the formal link between the boundary
function and the Loschmidt
echo~\cite{Heyl2013,Karrasch2013,Fagotti2013,Heyl2014} associated with a quantum
quench~\cite{Polkovnikov2011,Gambassi2012,Calabrese2011,*Essler2012}.
This formal link allows one to extend the Lee-Yang theory of phase transitions~\cite{Lee1952,*Yang1952} to nonequilibrium quantum dynamics.
Recently we studied~\cite{Hickey2013c} the connection between the singular
features in the FCS captured by these states and the emergence of temporal nonanalyticities in the return amplitude on quenching these states across the FCS critical line.
These temporal nonanalyticities dubbed dynamical phase transitions
(DPTs)~\cite{Heyl2013} often only appear for
quenches~\cite{Heyl2013,Fagotti2013,Hickey2013,Pollmann2010} in certain areas of
parameter space and in the previous studies quenching across either a quantum or FCS
critical point results in there emergence.  A recent work by Vajna et
al.~\cite{Vajna2013} highlighted that the presence of equilibrium phase
transitions does not imply DPTs will emerge, and so a natural question to
consider is whether this result applies to FCS transitions also.

The focus of this work is twofold, in the first part we focus on analytic
properties of the generating functions of the time-integrated transverse magnetization and anisotropy in the $1$d $XY$ model.  Examining the properties
of the special states associated with these observables, which we call the
$s$-states, we find the $2$nd order FCS phase transitions associated with these
observables correspond to $3$rd order static phase transitions in these states. 
Furthermore when considering the latter observable without any
anisotropy we find for \emph{every point in the ferromagnetic regime} there
exists a $3$rd order static phase transition. However contrary to expectation DPTs \emph{do not
emerge} when quenching across the FCS critical points of this observable.

In the second part of this study we
explore the preparation of these states in detail.  We study the likelyhood of
reaching the $s$-states using appropriate
Markovian baths and the timescales required to reach these states.  
We find that in the thermodynamic limit dissipation prevents one from preparing such a state
and one must consider an alternative method using an ancillary
system~\cite{Hush2013,Breuer2004,Breuer1999,Moodley2009,Imamog1994,Gambetta2002}.

We begin in Sec.~\ref{sec:TFIM} with a description of the theoretical framework
of time-integrated observables.  We then discuss the proposed method of
preparing such special states using Markovian baths in Sec.~\ref{sec:prep}.
Following this we present a theoretical primer on DPTs and the quench protocol
in Sec.~\ref{sec:DPT}.  Then in Sec.~\ref{sec:XYmodel} we present our results
on $3$rd order phase transitions and DPTs in the $XY$ model.  We discuss the
timescales required to prepare the $s$-state in Sec.~\ref{sec:timescales} along
with a discussion on the use of an ancilla system in preparing these states. 
To finish we present our conclusions in Sec.~\ref{sec:Conc}.

\section{Theoretical Background}

\subsection{Time-integrated observables}
\label{sec:TFIM}
A closed quantum system is described by its Hamiltonian, $H$, which defines the
time evolution of this system under an associated unitary operator.  To capture
the dynamics of this system we wish to examine the moments of a general
dynamical observable
\begin{equation}
Q_t = \int^{t} q(t') dt',
\end{equation}
here $q(t)$ is the operator of interest in the Heisenberg representation.  To
construct the MGF of this time-integrated observable we deform $H$ to a
non-Hermitian operator $H_s$.  Associated with this new operator is a
non-unitary evolution operator $T_t(s)$, both of these operators are defined as
\begin{equation}
T_t(s) \equiv {e}^{-itH_s}, H_s\equiv H-\frac{is}{2}q.
\end{equation}

From these definitions it is easy to that the generating function of $Q_t$ is
given by
\begin{equation}
Z_t(s) = \langle T_t^{\dagger}(s)T_t(s)\rangle.
\end{equation}
With the MGF the moments of $Q_t$ are generated simply via differentiation,
$\langle Q_t^n\rangle = (-1)^{n}\partial_s^n Z_t(s)|_{s\rightarrow 0}$, and the
CGF is simply the logarithm of this object, $\Theta_t(s)\equiv \log Z_t(s)$. 
These objects define the FCS~\cite{Nazarov2003}, but compared to the usual
approach where the FCS focusses on the characteristic function here we consider the generating function
with real $s$.  In the study of  FCS phase transitions it is useful to study
that analytic properties of a scaled version of the CGF in the long-time limit
\begin{equation}
\theta(s) = \lim_{N,t\rightarrow \infty} \frac{\Theta_t(s)}{Nt}.
\label{eq:SCGF}
\end{equation}

Making a connection with equilibrium thermodynamics we consider this CGF a full
dynamical ``free energy'' and define an order parameter,
$\kappa_s = -\partial_s\theta(s)$, and corresponding FCS susceptibility, $X_s
= \partial_s^{2}\theta(s)$.   Similar to their counterparts in equilibrium
statistical physics we use these quantities to characterize the FCS phases of the system and a diverging $X_s$ is indicative of a $2$nd
order FCS critical point.
Moreover for each point in parameter space we can define a state $\ket{s}$ which are right eigenstates of $H_s$ and play a
role analogous to the ground state of the $XY$ model.  Starting with an
initial state $\ket{i}$ this state is defined as $\ket{s} \equiv
\lim_{t\rightarrow \infty} T_t(s)\ket{i}$ with an appropriate normalization.
 With this state we define an $s$-biased expectation value of an observable 
\begin{equation}
\langle \mathcal{O} \rangle_s
\equiv \lim_{t\rightarrow \infty}
\frac{\bra{0}T_t^{\dagger}(s)\mathcal{O}T_t(s)\ket{0}}{Z_t(s)},
\end{equation}
where for the initial state $\ket{i}$ we use the ground state of the $XY$
model $\ket{0}$. Taking $\mathcal{O}$ to be the operator of interest $q$, in the
long-time limit it is easy to see that
\begin{equation}
\label{eq:constraint}
\frac{\langle q \rangle_s}{N} = -\frac{\theta(s)}{s}.
\end{equation}
From this equation it becomes apparent that the $2$nd derivative of the
$s$-biased expectation of $q$ with respect to $s$ will diverge at a FCS
critical point.
Therefore a $2$nd order FCS transition is equivalent to a a $3$rd order phase
transition exists in this extended $s$-state regime.

\subsection{Preparation of $s$-state}
\label{sec:prep}
The $s$-state is the result of long-time evolution under a non-unitarty
evolution operator~\cite{Hickey2013}, this operator evolves the density matrix
$\rho(t)$ according to $\dot{\rho}(t) = -i[H,\rho(t)]-\frac{s}{2}\{q,\rho(s)\}$. This evolution is
similar in appearance to a Lindblad master equation without recycling terms,
$\dot{\rho}(s) = -i[H,\rho(s)] + \sum_i (L_i\rho L_i^{\dagger}
-\frac{1}{2}\{L_i^{\dagger}L_i,\rho\})$, this full evolution describes a
system connected to a Markovian bath.   Noticing that if the observable of
interest $q$ is related to the jump operators $L_i$ via $\sum_i
L_i^{\dagger}L_i = sq$ then $T_t(s)$ defines this associated open quantum system
in between quantum jumps~\cite{Plenio1998,Gardiner2004}.  From this the MGF is
simply the probability of no jumps occurring $P_0(t)$ up to a time $t$ in the associated open quantum system
and $s$ plays the role of the decay rate.

For the case of the TFIM and the total transverse magnetization,
if we make a trivial shift so that $q = \sum_i (\sigma_i^z + 1)$, we can define
the jump operators as $L_i = \sqrt{2s}\ket{-}_i\bra{+}_i$ where
$\sigma_i^z\ket{\pm} = \pm\ket{\pm}$.  It was shown that using cold ions one could simulate this
associated open quantum system and from the jump statistics extract features
of the critical curve for finite sizes and short times.  The question we pose
now is this, if we have access to such baths can we prepare the $\ket{s}$ state
directly from the no jump evolution and tune the decay rate to probe this
critical line? Furthermore, if this is possible what is the timescale required to prepare these
$s$-states?

\subsection{Return Amplitude and Quantum Quenches}
\label{sec:DPT}
One of the central quantities in equilibrium statistical physics is the boundary
partition function
\begin{equation}
 Z(L) = \bra{\psi_a}{e}^{-L H}\ket{\psi_b},
\end{equation}
where $L$ is the length of the boundary, $H$ is the Hamiltonian and
$\ket{\psi_{a/b}}$ are the boundary states.  This quantity may be contected to
the nonequilibrium protocol known as the quantum quench by analytically
continuing to the complex plane, $L \rightarrow \beta$ where $\beta \in \Com$. 
Taking identical boundaries $a = b = 0$ one can readily show that if $\beta = i
t$ the analytically continued boundary partition function is the Loschmidt
amplitude
\begin{equation}
G(t) =  \bra{\psi_0}{e}^{-iHt}\ket{\psi_0} \,.
\end{equation}

In the context of the quantum quench $\ket{\psi_0}$ is the initial state and $H$
is the post quench Hamiltonian.  The boundary partition function has zeros in
the complex $L$ plane and in the thermodynamic limit these zeros may coalesce to
form a critical line~\cite{Fisher1965}. If this line intersects
the real $L$ axis for some parameter values then the system will under a phase transition.  However this
critical line may also intersect the imaginary $L$ axis and result in so-called
dynamical phase transitions~\cite{Heyl2013}.  Analogous to equilibrium
statistical physics these DPTs manifest as temporal nonanalyticities in the
large deviation function~\cite{Touchette2009} associated with the return
amplitude

\begin{equation}
l(t) =\lim_{N\rightarrow \infty} \frac{-1}{N}\log |G(t){|}^{2},
\end{equation}
where here $N$ represents the system size.  It is important to note that
although this mapping is formally exact the return amplitude does not
provide information on the equilibrium statistics in this problem. 

In this work we consider a specific type of quantum quench protocol which we
refer to as the $s$-quench~\cite{Hickey2013c}.  The protocol is as follows: the
system is initially prepared in the relevant $\ket{s}$ state, then it is ``quenched'' to $s =
0$ by allowing the state to evolve under the original $s = 0$ $XY$ model
Hamiltonian $H$,
\begin{equation}
\ket{s_t} = {e}^{-itH}\ket{s}.
\end{equation}
In the next Section we will examine the analytic properties of the return
amplitude associated with this ``quench'' protocol.

\section{$3$rd order phase transitions in the XY model}
\label{sec:XYmodel}
\begin{figure}[h!]
\includegraphics[width = 1.0\columnwidth]{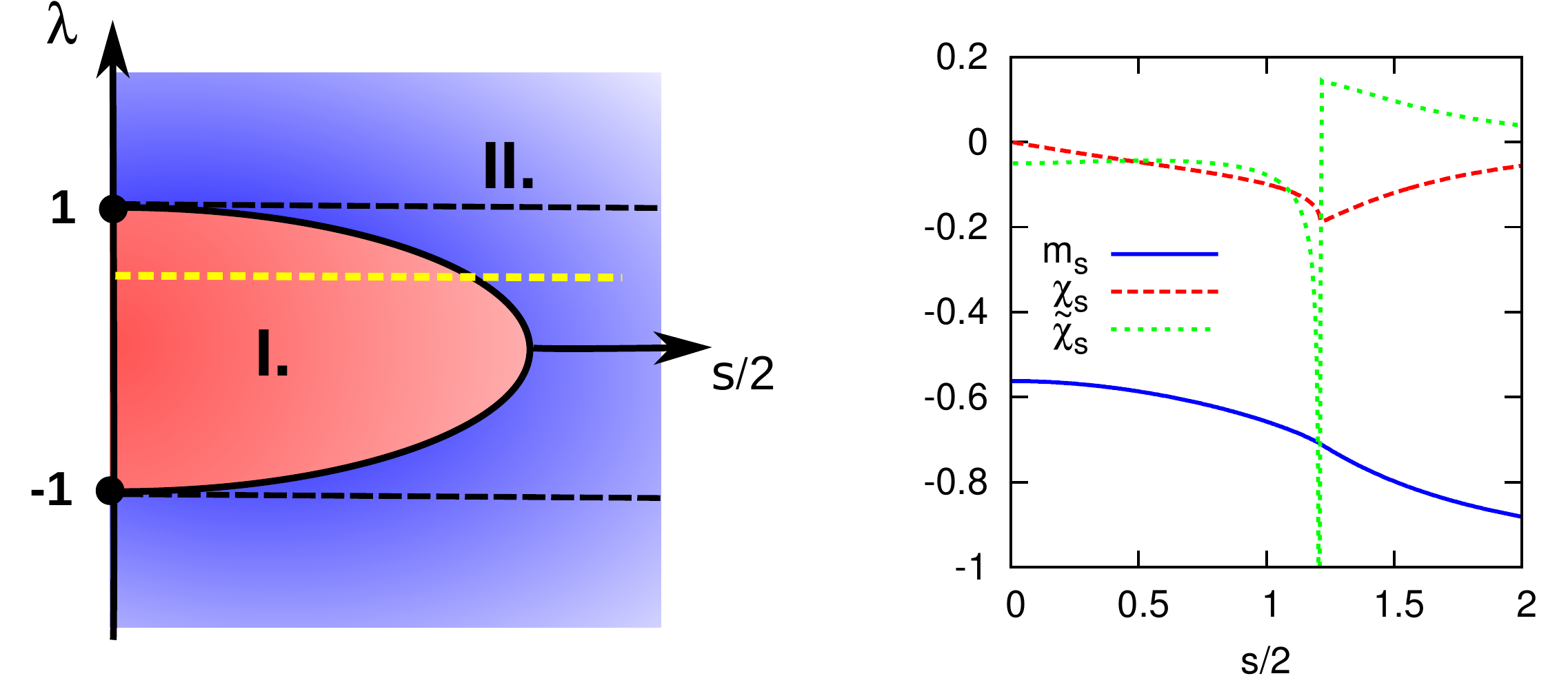}
\caption{(Color Online)  The left panel is the FCS phase diagram of the XY
model with finite anisotropy parameter.  The right panel shows the
magnetization in the $s$-state along the $\lambda = 0.5 ~(\gamma = 1.4)$.  The
magnetic susceptibility exhibits a ``kink'' at the phase boundary between
regions I and II.  This leads its derivative $\tilde{\chi}_s$ to diverge at the
phase boundary indicating a $3$rd order phase transtion.}\label{fig:fig1}
\end{figure}
We now focus on the $1$d $XY$ model with periodic boundary conditions, this is a
paradagmatic model of a quantum phase transition and is defined by the
Hamiltonian
\begin{align}
H &= -\sum_i \frac{1+\gamma}{2}\sigma_i^x\sigma_{i+1}^x-\sum_i
\frac{1-\gamma}{2}\sigma_i^y\sigma_{i+1}^y - \lambda\sum_i \sigma_i^z
\nonumber
\\
&= \sum_k \epsilon_k(\lambda)(A_k^{\dagger}A_k - 1/2),
\end{align}
where $\sigma^{x,z,y}$ are pauli spin operators, $\gamma$ is the anisotropy
parameter and $\lambda$ denotes the strength of the transverse field.  Using a
Jordan-Wigner transformation followed by a Bogoliubov rotation this Hamiltonian may be diagonalized and has an energy
spectrum
\begin{equation}
\epsilon_k(\lambda) = 2\sqrt{(\lambda-\cos k)^{2} +\gamma^2\sin^{2}k}.
\end{equation}
This model has critical points at $\lambda = \pm 1$, where the ground state
changes from a ferromagnetic state to being paramagnetic in a singular fashion. 
Within the ferromagnetic regime there is also a critical line along $\gamma =
0$ either side of which the ground state is ferromagnetic with the spins aligned along either the $x$ or $y$
direction.  

We consider time integrals of the transverse
magnetization and the total anisotropy in the ground state of this model. 
Beginning with the former we set $q = \sum_{i} \sigma_i^{z}$, the deformed
$H_s$ is still diagonalizable via free fermion methods and the analytic form of the CGF is
accessible,
\begin{equation}
\theta(s) = \frac{2}{\pi}{\text I}{\text m}\Bigl(\int_0^\pi dk |\sqrt{(\lambda
+ is/2 -\cos k)^{2} + \gamma^2\sin^{2} k}|\Bigr).
\end{equation}

The dynamic susceptibility associated with this diverges along a critical curve
in the $s$-$\lambda$ plane, of which the end points are the static quantum phase
transtions.  This curve obeys the equation
\begin{equation}
\lambda^2 +(s/2\gamma)^{2} = 1, ~|\lambda| \le 1,
\end{equation}
where for each value of $\lambda$ there is a critical $k$ mode whose associated
energy spectrum becomes $0$ at the critical line, $k^* = \cos^{-1} \lambda$.
We note that taking $\gamma = 0$ we obtain the critical curve of
Ref.~\cite{Hickey2013}.
This curve marks a $3$rd order phase transition in the magnetization of the $\ket{s}$
states, where the static susceptibility $\chi_s = \partial_s \langle m_x
\rangle_s$ has a ``kink'' and its derivative $\tilde{\chi}_s$ diverges, this is
shown in Fig.~\ref{fig:fig1}. 

\begin{center}
\begin{figure}
\includegraphics[width = 1.0\columnwidth]{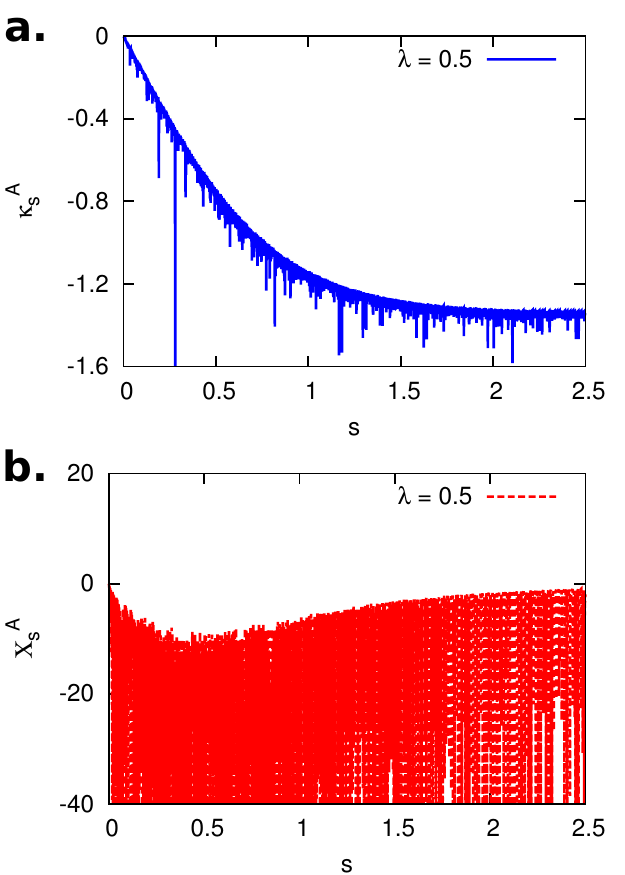}
\caption{(Color Online) (a) The dynamical activity of the
anisotropy parameter for $N = 2000$ spins  is plotted, tuning $s$ many
discontuities associated with FCS singularities can be seen.  (b) The
discontinuities in $\kappa^A_s$, see above, manifest as divergences in the
dynamical susceptibility and mark a $3$rd order static phase transition in the associated
$\ket{s}$-states.}\label{fig:fig2}
\end{figure}
\end{center}

Switiching focus to the time-integrals of the anisotropy parameter $q =
\sum_i \sigma^x_i \sigma^x_{i+1}-\sigma^y_i \sigma^y_{i+1}$, we denote the relevant
quantities of interest with a superscript $A$.  The relevant $H^{A}_s$ is once
again diagonalizable using free fermion techniques~\cite{Sachdev2011} and we
find the FCS ``free'' energy is 
\begin{equation}
\label{eq:atheta}
\theta^{A}(s) = \frac{2}{\pi}{\text I}{\text m}\Bigl(\int_0^\pi dk
|\sqrt{(\lambda -\cos k)^{2} + (\gamma+is)^2\sin^{2} k}|\Bigr).
\end{equation}

When $\gamma \neq 0$ there are no FCS critical points but along the
anisotropy coexistence line $\gamma = 0$ the susceptibility $X^A_s$ diverges
with a square root singularity when $\lambda = \cos k \pm {s} \sin k$. 
Examining this equation we find the critical $k$ mode which satisfies this
equality is given by
\begin{equation}
\label{eq:critk}
k^{*} = |\cos^{-1}\Bigl[ \frac{\lambda\pm
s\sqrt{1+s^2-\lambda^2}}{1+s^2}\Bigr]|,
\end{equation}
with the constraint that the $k$ modes lie in the range $[0,\pi]$. 
Despite this restriction on the range of $k$ modes we find that in ferromagnetic regime $|\lambda|< 1$ there are
two solutions to Eq.~\eqref{eq:critk}. Therefore
this region is critical as there are $3$rd order phase transitions
in the $\ket{s}$ states for every point in the $s$-$\lambda$ region.  Moving to
the paramagnetic regime $ -1 > \lambda >1$ solutions to Eq.~\eqref{eq:critk}
only exist for this equation when $1+s^2 \ge \lambda^2$.  This results in
the emergence of $1$st order FCS transition at $s_c = \pm\sqrt{\lambda^2-1}$,
beyond this critical point there once again exists solutions to Eq.~\eqref{eq:critk} and
$3$rd order FCS phase transitions at each point in parameter space.  

\begin{figure}
\includegraphics[width = 1.0\columnwidth]{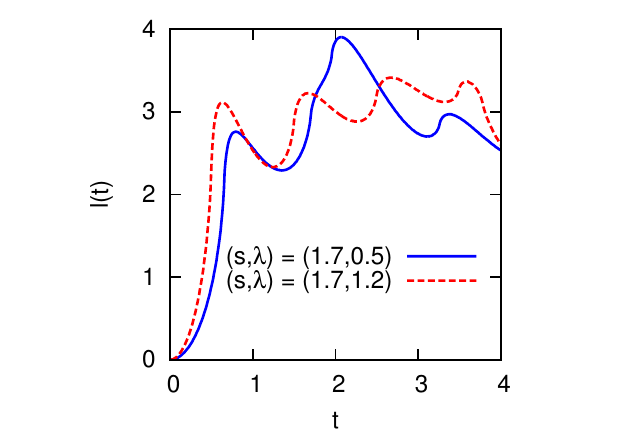}
\caption{(Color Online) Quenching from points $(s,\lambda)$ to $(0,\lambda)$
there is no evidence of DPTs despite crossing numerous FCS
transition points.  This highlights that the crossing a FCS critical point
under a nonequilibrium protocol is not a sufficient criteria
for the emergence of DPTs.}\label{fig:fig3}
\end{figure}

When there exists no $k^{*}$ the spectrum of $H^{A}_s$ is real and there
exists an associated unbroken
$\Pa\T$-symmetry~\cite{Bender1998,Bender2007,Mostafazadeh2010}, for a regular
Hermitian operator this symmetry is simply Hermitian conjugation.
We do not discuss the technical details of this symmetry but it has implications
on the temporal scaling of the cumulants.  This is easy to see from Eqs.~\eqref{eq:SCGF} and ~\eqref{eq:atheta}, in the
ferromagnetic regime the spectrum of $H_s$ has complex eigenvalues which contribute to
$\theta^{A}(s)$.  From Eq.~\eqref{eq:SCGF} this implies the cumulants at
$s = 0$ scale at least linearly in time, however in the paramagnetic regime
$\theta^{A}(s)$ is strictly $0$ in the vicinity of  $s = 0$. This is because
for $|s| < |\lambda|$ the eigenvalues of $H^{A}_s$ are real, the form of the CGF
in this parameter regime implies the physical cumulants are oscillatory or scale
sublinearly with time.
These results are interesting and highlight the importance of the spectrum of
$H^{A}_s$ when considering time-integrated observables, as
discussed in Ref.~\cite{Hickey2014}.

Starting from the critical regime~($|\lambda| < 1$), the analytic form of
$\ket{s}$ is analytically accessible (see Appendix).  Quenching this state to
$s = 0$ we find the return amplitude rate function takes the form
\begin{equation}
\label{eq:DPT}
l(t) = -2{\text R}{\text e}\Bigl(\int^{k_2}_{k_1}
\frac{|\sin\alpha^s_k|^{2}+|\cos
\alpha^{s}|^{2}{e}^{-2it\epsilon_{k}(\lambda)})}{\cosh(2{\text I}{\text
m}(\alpha^{s}_{k}))}\Bigr),
\end{equation}
where $\alpha^s_k$ is the difference between Bogoliubov angles
which arise in diagonalizing $H^{A}_s$ and $k_{1,2}$ are the upper and lower
solutions to Eq.~\eqref{eq:critk} respectively.  Quenching from both within the
critical regime $|\lambda|< 1$ and across the $1$st order FCS critical
line~($|\lambda|>1$) does not result in the emergence of DPTs, this is shown in
Fig.~\ref{fig:fig3}.
Although for finite $N$ there exists two families of solutions, $k = k_{1,2}$,
where $|\cos \alpha^{s}_{k}|=|\sin \alpha^{s}_{k}|$, in the thermodynamic limit
these zeros do not result in DPTs emerging in Eq.~\eqref{eq:DPT}.  The lack
of DPTs on quenching across these static critical points in $\ket{s}$ space is surprising and highlights that the presence of such critical points is not
sufficient for these DPTs to emerge upon quenching.

\section{Timescales to reach $s$-state}
\label{sec:timescales}
\begin{figure*}
\includegraphics[width = 1.7\columnwidth]{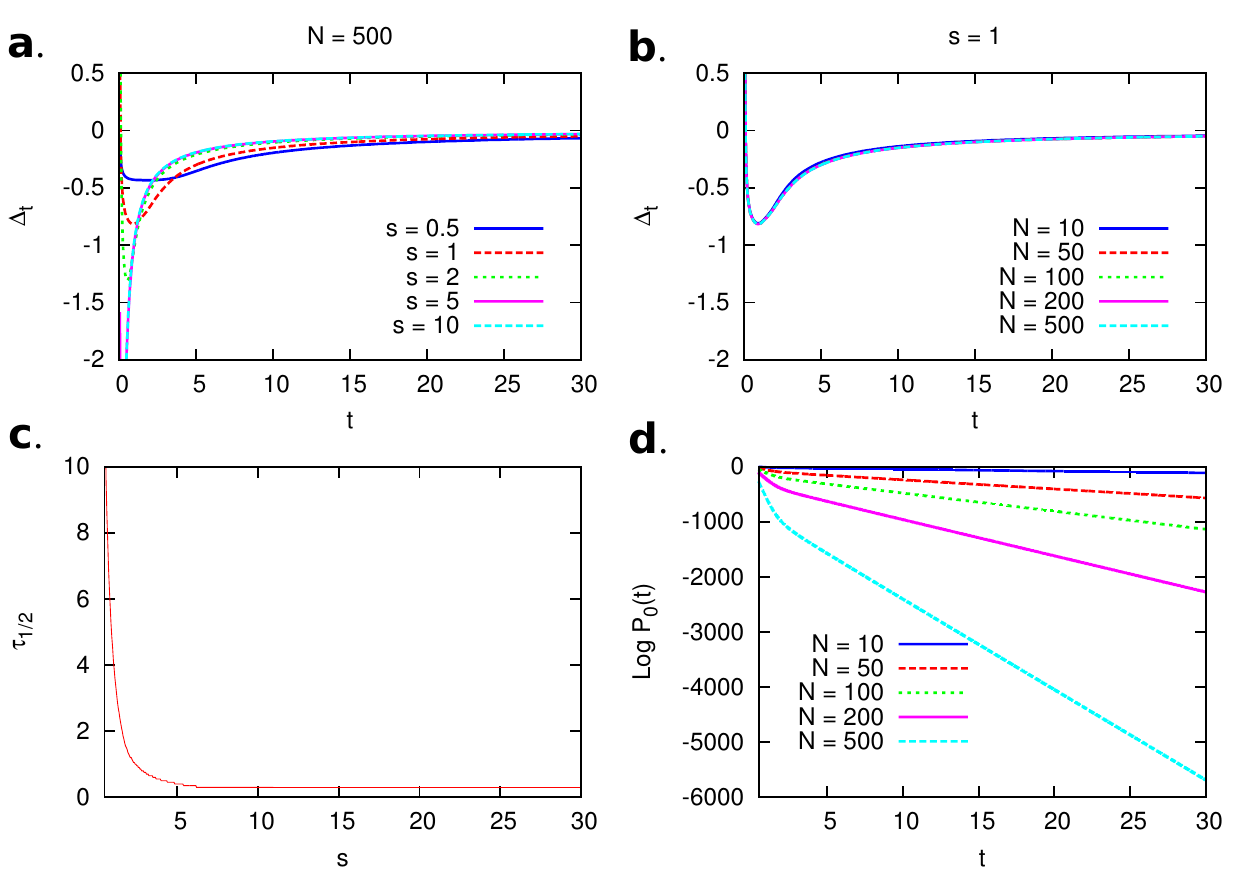}
\caption{(Color Online) (a,c) The dynamical evolution towards $\ket{s}$ is
sensitive to decay rate for states with $s\sim\OO(1)$ but on increasing $s$ the
magnitude of the decay rate has little impact in the evolution towards $s$.
(b,d) The timescale for the non-unitary evolution is independent of the system
size, however the probability of the system undergoing this evolution where no
jumps occur tends to $0$ in the thermodynamic limit.}\label{fig:fig4}
\end{figure*}
Having discussed the $3$rd phase transitions that may occur in this model we now
focus on the timescales required to prepare the $\ket{s}$ states associated with
the time-integrated transverse magnetization in the TFIM, therefore $\gamma =
1$ throughout this section.  Specifically we are interested in the $\ket{s}$
states which undergo $3$rd order phase transitions by tuning $s$ and so we
take $\lambda = 0.5$ in this section. We begin by examining the timescale
required for $Z_t(s)$ to converge to ${e}^{tN\theta(s)}$ as a function of the number of spins $N$, and the decay rate $s$.  We measure this convergence by examining the scaled ``distance'' to the long-time regime

\begin{equation}
\Delta_t = \frac{\log Z_t(s)/Nt - \theta^{N}(s)}{\theta^{N}(s)},
\end{equation}
where $\theta^{N}(s)$ is the CGF of Eq.~\eqref{eq:SCGF} for finite $N$.  When
$\Delta_t = 0$ the ground state has converged to $\ket{s}$ under the no jump evolution of the bath.  Taking $N = 500$ we find to
strictly reach the $\ket{s}$ state requires an infinitely long-time for all
values of $s$.  However the timescale required to produce a state close to $\ket{s}$, $|\Delta_t|
< 0.05$,  decreases with increasing $s$.  This is captured by the timescale
$\tau_{1/2}$ defined by $\Delta_{\tau_{1/2}} = 0.5$, where this equation may
have multiple solutions so we take the largest $\tau_{1/2}$.  This timescale is
plotted in Fig.~\ref{fig:fig4}(c) and shows that as $s\rightarrow 0$ the
timescale required to prepare a state very close to the $\ket{s}$ state
diverges.  This timescale decreases with increasing $s$ but the rate of decrease
with increasing decay rate tails off for $s \gg 1$.  Focussing on fixed $s$ and
varying system size we find the timescale required to reach the final state is
independent of the system size, this is shown in Fig.~\ref{fig:fig4}(b).  

The evolution of the system when connected to these Markovian environments is
stochastic in nature and so there is only a finite \emph{probability} that the system will evolve under the no jump evolution
defined by $H_s$.  This probability is known as the \emph{survival probability}
and is analytically calculable for this model.  In fact due to the shift in
observable to $q =\sum_{i}(\sigma^{z}_i + 1)$ the survival probability is
trivially related to the MGF of the time-integrated transverse magnetization by the
equation
\begin{equation}
P_0(t) = {e}^{-sNt}Z_{t}(s).
\end{equation}

This quantity is highly dependent on system size and in the thermodynamic limit
is zero for all finite $t$.  This result is shown in Fig.~\ref{fig:fig4}. 
Combined with the system size independence of $\Delta_t$ we conclude that to
prepare a state close to $\ket{s}$ it is better to use a \emph{few body system}.
Such few body open quantum systems may be experimentally probed via digital
simulation with ultracold ions~\cite{Barreiro2011,Blatt2012,Muller2011}. 
Although the finite size effects for such small system sizes are large, it was demonstrated in Ref.~\cite{Hickey2013}
that some signatures of the transition are still extactable.

If one wishes to probe the $3$rd order phase transition in the thermodynamic
limit it is necessary to evolve the system under the no jump evolution directly and avoid
the prospect of emissions which prevent us from preparing the system close to
the $\ket{s}$ state.  This may be done via introducing an ancilliary two level system in
conjunction with the Markovian baths. Consider the evolution of the full
system$+$ancilla qubit in a Markovian environment, its density matrix $\varrho$
evolves under a master equation
\begin{equation}
\dot{\varrho} = -i[\mathcal{H},\varrho] + \sum_i \mathcal{J}_i\varrho
\mathcal{J}^{\dagger}_i -
\frac{1}{2}\{\mathcal{J}^{\dagger}_i\mathcal{J}_i,\varrho\}.
\end{equation}

Choosing $\mathcal{H} = H\otimes \ket{0}\bra{0}$, $\mathcal{J}_{i} = L_i \otimes
\ket{1}\bra{0}$, where $H$ is the TFIM Hamiltonian and the projectors
$\ket{0}\bra{0}$ and $\ket{1}\bra{0}$ act on the ancilla subspace, one can
readily show that the no jump evolution defined by $H_s$ is readily given by 
\begin{align}
\dot{\rho} &= \Tr_{Anc.}(\mathbb{I}\otimes\ket{0}\bra{0}\dot{\varrho}) \\
&= -i[H,\rho] - \frac{1}{2}\{L^{\dagger}_iL_i,\rho\}, \nonumber
\end{align}
where the trace is performed over the ancilla subspace.  Using this simple two
level ancilliary system one may implement the no jump evolution required to
prepare a state very close to $\ket{s}$ in the thermodynamic limit and thus
probe the $3$rd order static phase transitions in this extended state space.

\section{Conclusions}
\label{sec:Conc}
In this paper we examined the singularities of the generating functions of the
time-integrated transverse magnetization and anisotropy in the $1$d $XY$ model. 
In the former case there exists an elliptical curve of FCS critical points in
the $s$-$\lambda$ plane where the eccentricity of the ellipse is set by the
anisotropy $\gamma$.  In the latter case when $\gamma = 0$ we
uncovered a critical regime where every point in parameter space has an FCS singularity
but the FCS singularities of the anisotropy are not marked by the emergence
of DPTs in the return amplitude upon quenching.
The FCS singularities in both cases manifest themselves as $3$rd order phase
transitions in the $\ket{s}$ states. We also examined the timescales required to
prepare the $s$-states which exhibit these novel phase transitions.  We focussed
on the example of the TFIM and the $s$-states associated with the time-integrated transverse magnetization.  We found that,
starting in the ground state, the timescale to reach the $s$-state is
independent of the system size and is weakly dependent on the decay rate when
$s \gg 1$. However the probability of evolving without emission becomes zero in
the limit of large system size due to dissipation. Thus to prepare
$\ket{s}$ and find features of the $3$rd order phase transitions few body
systems are preferrable such as cold ion systems used in digital simulation. 
To reach the thermodynamic limit one needs to use an ancilliary system where the desired evolution is contained as a
block of the joint system$+$ancilla density matrix.  Beyond their properties
in capturing FCS singularities as static quantum critical points and DPTs, many
questions remain about the nature of these $\ket{s}$ states.  It would be
interesting to study the entanglement properties of the $\ket{s}$ states and
their relationship to scaling theories in non-unitary conformal field theories.  Finally there is the interesting
question of how static observables in the ground state relate to the properties
of these $\ket{s}$ states.

\section{Acknowledgements}
We thank Sam Genway for useful comments on the manuscript and fruitful
discussions.  This work was supported by the European Union Research Scholarship
provided by the University of Nottingham.

\appendix

\section{Diagonalizing $H^{A}_s$ and finding $\ket{s}$}
The non-Hermitian operator associated with the time-integrated anisotropy is
given by
\begin{equation}
H^{A}_{s}= -\sum_i \frac{1+\gamma+is}{2}\sigma_i^x\sigma_{i+1}^x-\sum_i
\frac{1-\gamma-is}{2}\sigma_i^y\sigma_{i+1}^y - \lambda\sum_i \sigma_i^z.
\end{equation}
To diagonalize this operator we first perform a Jordan-Wigner transformation
followed by Bogoliubov rotation~\cite{Sachdev2011}.  The first transformation
maps the spin operators at each site $\sigma^z_i$,$\sigma^+_i$, and $\sigma^-_i$ to fermionic
creation and annihilation operators $c_i$ and $c^{\dagger}_i$ with
$\{c^{\dagger}_i,c_j\} = \delta_{i,j}$ via
\begin{equation}
\label{eq:JWT}
\begin{split}
\sigma^{z}_{i} &= 1-2c^{\dagger}_{i}c_{i},\\
\sigma^{+}_{i} &= \prod_{j<i}(1-2c^{\dagger}_{j}c_{j}) c_{i}, \\
\sigma^{-}_{i} &= \prod_{j<i}(1-2c^{\dagger}_{j}c_{j}) c^{\dagger}_{i}.
\end{split}
\end{equation}
Fourier transforming the transformed Hamiltonian results in an operator which
may be diagonalized via Bogoliubov rotation
\begin{equation}\label{eq:BogRot}
\begin{split}
c_{k} &= \cos \frac{\phi^{s}_{k}}{2}~B_{k} + i\sin \frac{\phi^{s}_{k}}{2}
~\bar{B}_{-k},\\
c^{\dag}_{k} &= \cos \frac{\phi^{s}_{k}}{2}~\bar{B}_{k} -
i\sin \frac{\phi^{s}_{k}}{2}~B_{-k}.
\end{split}
\end{equation}
It is worth noting here we restrict ourselves to an even number of spins $N$ and
assume periodic boundary conditions.  With this rotation $H^{A}_s$ is diagonal
in the complex fermionic pair  $\{ \bar{B}_{k'},B_{k}\} =
\delta_{k',k}$, where $\bar{B}_{k} \neq B^{\dagger}_{k}$ provided
$\phi^{s}_{-k}=-\phi^{s}_{k}$ and 
\begin{equation}
\tan \phi^{s}_{k} = \frac{(\gamma+is)\sin k}{\lambda-\cos k }.
\end{equation}
In the case of no anisotropy~($\gamma = 0$) the free fermion dispersion of
$H^{A}_s$ is then found to be
\begin{equation}
\epsilon_k(\lambda,s) = 2\sqrt{(\lambda - \cos k)^{2} -s^2\sin^{2}k}.
\end{equation}

The key property in determining the evolution of the ground state $\ket{0}$ of
$H^{A}_{s = 0}$ is that the fermionic states at $s = 0$ may be expressed in
terms of the fermionic states at finite $s$.  In fact the ground state is simply
a BCS state of $H^{A}_{s}$
\begin{align}
\label{eq:BCS}
\ket{0}&= \frac{1}{\mathcal{N}'} \exp\left(\sum_{k>0}
B(k)\bar{B}_{k}\bar{B}_{-k}\right)\ket{0}_{s} \nonumber \\
&=\frac{1}{\sqrt{\cosh(2{\text I}{\text m}(\alpha^s_k))}} \bigotimes_{k>0}
\left[ \cos{{\alpha}_{k}^{s}}{\ket{0_{k},0_{-k}}}_{s} -i\sin{{\alpha}_{k}^{s}}{\ket{1_k,1_{-k}}}_{s} \right]
\end{align}
Here the $k$th mode of the $s$-vacuum $\ket{0_{k},0_{-k}}_{s}$ is defined such
that $B_{\pm k}\ket{0_{k},0_{-k}}_{s} = 0$ and the complex angles in the
coefficients is simply $\alpha^s_k = \frac{\phi^{s=0}_k - \phi^s_k}{2}$.  While
$\ket{1_k,1_{-k}}_s = \bar{B}_k\bar{B}_{-k}\ket{0_{k},0_{-k}}_{s}$ signifies the
occupied fermionic state with the wavevector $|k|$ that diagonalizes
$H^{A}_{s}$. With Eq.~\eqref{eq:BCS} we can evolve $\ket{0}$ under the non-unitary evolution defined by $H^{A}_s$ to obtain the $\ket{s}$.  There is however a subtlety here, as often there exists some $k$ modes for which $\epsilon^{A}_k(\lambda,s)$ is
real.
The $\ket{s}$ state is then chosen such that Eq.~\eqref{eq:constraint} holds
and from \eqref{eq:SCGF} it is clear in the long-time limit these $k$ modes do
not contribute to $\theta^{A}(s)$ and hence $\ket{s}$.
Thus we find 
\begin{equation}
\ket{s} = \prod_{k>0,~ {\text I}{\text m} \epsilon^{A}_k\neq 0}
\frac{\ket{1_k,1_{-k}}_s}{\sqrt{\cosh (2{\text I}{\text m}(\alpha^s_k))}}.
\end{equation}

\end{document}